\documentclass[epsfig,prd,aps,preprint,superscriptaddress,showkeys]{revtex4}
\usepackage{graphics}
\usepackage{natbib}

\begin{document}

\title{Equipartition of Energy in Gravitating Systems}

\author{Diego Pav\'on}
\email{diego.pavon@uab.es} \affiliation{Department of Physics, Faculty of Sciences, Autonomous University of Barcelona,\\ 08193 Bellaterra (Barcelona) Spain.}

\begin{abstract}
\noindent  We explore on the one hand, whether the well-known theorem of equipartition of energy also applies to physical systems in which gravity plays a non-negligible role, as is the case of cosmological horizons and, on the other hand, if the holographic principle holds for these systems. Furthermore, we find a direct connection between the partition function of the gravitating system and the metric of the gravitational field. 
\end{abstract}
\[\]

\maketitle
\section{Introduction}
\noindent Unlike to what it have looked like for so many years, gravitation and thermodynamics are no longer two separate disciplines. Indeed, starting with the pioneering work of Tolman (see, e.g. \cite{Tolman1934}) both branches of Physics have become more and more interwined, especially after the realization that black holes are thermodynamic objects \cite{Hawking1975, Hawking1976} and that cosmological horizons are endowed with thermal properties \cite{Gibbons1977}. Here we focus on the well-known equipartition theorem which immediately follows from the principles of statistical mechanics and applies to many aspects of the non-gravitational world. Since we know now that gravitation and thermodynamics are more closely related that we first thought, it is natural to ask whether its validity also extends to systems not dominated by the electromagnetic field but by gravity, as it is the case of cosmological horizons. The main  purpose of this short paper is to answer this question. Curiously enough, in spite of the subject of black hole thermodynamics being about fifty years old, to the best of our knowledge, this question has not been considered so far. 
\\  

\noindent According to the equipartition  theorem, the energy of a macroscopic system in thermal equilibrium is proportional to the product of the number of unfrozen degrees of freedom, ${\cal N}$,  by its temperature $T$,
\begin{equation}
{\cal{E}} = \zeta \, \, {\cal{N}} k_{B} \, T,
\label{eq:eqp1}
\end{equation}
(see e.g. \cite{Huang,Pathria}) where $\zeta$ is a number of order unity.  As a consequence, every particle of the system contributes an equal amount, namely 
$\zeta  k_{B}  T$, to the total energy of the latter.  

\noindent The theorem was established for systems dominated by the electromagnetic interaction. Whether it is also valid for systems dominated by gravity is debatable. The case of black holes illustrates this; while it holds for Schwarzschild and Kerr black holes it seems not to  for charged black holes \cite{dpavon}. An isothermal column of gas immersed in a gravitational field whose potential, $\phi$, depends on the height constitutes another example. In this situation heat flows from points at a higher potential to points at a lower potential  \cite{Eckart1940,pla1980}. For the system to be in thermal equilibrium, i.e. so that the heat flow vanishes, the temperature must vary with height according to $T(1 \,+ \, c^{-2} \phi) =$ constant. Hence, if the system is at equilibrium,  particles at different heights will contribute to the total with distinct amounts of thermal energy.
\\

\noindent Our aim is twofold. (i)  To study if this theorem also holds for cosmological horizons of homogeneous and isotropic universes; and (ii) to explore whether the holographic principle \cite{Gerard,Susskind} is uphold (for an ample discussion of this principle see Bousso \cite{Bousso2002}). Thus far, the latter is not very precisely established. Loosely speaking, it asserts that the physics in a region limited by a causal horizon surface gets reflected on that surface. Here, we  shall content ourselves by seeing whether the value of the energy of the region enclosed by the causal surface coincides with the value of the energy associated to that surface. If they coincide, it seems reasonable to think that, at least up to some extent, the holographic principle is satisfied. In addition to that, we find a relationship between the metric of the gravitational field and the partition function of the system.
\\

\noindent Before going any further, it is worthwhile to recall that the number of degrees of freedom of the horizon and the entropy are usually assumed to be proportional to the area of the horizon (more specifically, ${\cal N} = {\cal A}/\ell_{pl}^{2}$, where $\ell_{pl}$ is Planck's length and $S = k_{B} {\cal A}/4$ the entropy).  Hence, one may think that Euler's relation, ${\cal{E}} = T S$ \cite{Callen1962} automatically justifies the theorem. However, this is not true because the validity of aforesaid relation, which  rests on the assumption that the entropy is a homogeneous first-order function of energy, fails in the case of both, black holes and cosmological horizons. So, we shall not resort to Euler's relation.
\\

\noindent The plan of this paper is as follows. First, we consider the event horizon of the De Sitter universe. Secondly, the Hubble and event horizons of universes whose scale  factor obeys a power-time law. Thirdly, we shall be concerned with phantom dominated universes whose factor of scale is of pole-like type. Finally, in the fourth section, we summarize our results and add some comments. We left aside particle horizons because, as far as we know,  no temperature has been assigned to them. In our units $c= k_{B} = G = \hbar =1$.

\section{De Sitter horizon}
\noindent The metric of a homogeneous and isotropic empty universe dominated by a real-valued parameter, called the cosmological constant $\Lambda$, can be written as

\begin{equation}
ds^{2} = - dt^{2} \, + \, [\exp{(2 (\Lambda/3)^{1/2} \, t)}] \, [dr^{2} \,+ \,  r^{2} d \Omega^{2}].
    \label{DSmetric}
\end{equation}

\noindent Consequently, the radius of the event (De Sitter) horizon and its area are  

\begin{equation}
r_{H} = a(t) \, \int_{t}^{\infty}{dx/a(x)} = \sqrt{3/\Lambda}
\label{radiusDeSitter}
\end{equation}
and
\begin{equation}
{\cal A} = 4 \pi r_{H}^{2} = 4 \pi \, \frac{3}{\Lambda},
\label{DSarea}
\end{equation}
respectively.
\\   

\noindent We assume that the event horizon is at equilibrium with itself and resort to Gibbs equation $d{\cal E} = T \, dS$, where $T = \sqrt{\Lambda/3}/2 \pi$ \cite{G&H} is the horizon's temperature and $S = {\cal{A}}/4$. Notice that the usual pressure term in that equation vanishes when this equation is applied  to any horizon because the latter is just a geometrical entity (not a material entity) and, as such, it cannot feel the ambience pressure (or any other pressure, for that matter). Therefore,
\\  

\begin{equation}
d{\cal E} = \frac{1}{2 \pi} \sqrt{\frac{\Lambda}{3}} \, d\left(\pi \, \frac{3}{\Lambda} \right). 
\label{TdS}
\end{equation}

\noindent Upon integration
\begin{equation}
{\cal E}  =  - \frac{\sqrt{3}}{2} \, \int_{-\infty}^{\Lambda}{ x^{-3/2} \, dx} 
\label{upon-integration1}
\end{equation}
we get the energy associated to the horizon,

\begin{equation}
{\cal E} = \sqrt{3/\Lambda}.
  \label{horizon-energy}
\end{equation}
\\  

\noindent Owing to the fact that the parameter $\Lambda$ may take any real value we carried out the integration between minus infinity and $\Lambda$. Clearly, last equation tells us that the latter cannot be negative. 
\\

\noindent We are now in position to check if the equipartition theorem coincides with 
the above expression for ${\cal E}$. As said before, we assume that the number of unfrozen 
degrees of freedom of the horizon is given by its area over the Planck's length 
to the square. In our units,  ${\cal N} = 12 \pi/\Lambda$. So, $\zeta {\cal N} T$ is proportional 
to ${\cal{E}}$ and reduces to it for  $\zeta = 1/2$. It is worth mentioning that the 
equipartition theorem is satisfied despite the fact that the heat capacity of 
the event horizon, $d{\cal E}/dT$, is negative. 
\\   

\noindent The physical meaning of ${\cal E}$ is not well understood. Let's explore 
its connection, if any, to the energy inside the volume bounded by the horizon. This is

\begin{equation}
E = \frac{4 \pi}{3} \, r_{H}^{3} \, \rho_{\Lambda} =  \frac{1}{2}\, \sqrt{\frac{3}{\Lambda}}.
 \label{Einside}
\end{equation}
\noindent Thus, $E =  \textstyle{1 \over 2} {\cal E}$.  The fact that both energies differ from each other, 
though just by a multiplicative numerical factor, tells us that the holographic principle is not fully satisfied. (According to one particular version of the said principle, the number of degrees of freedom contained in the volume bounded by the horizon (the ``bulk") must coincide with the number of degrees of freedom on the horizon's surface \cite{Paddy}). 
This should not come as a surprise since the entropy of the bulk vanishes (recall that $p_{\Lambda} = - \rho_{\Lambda}$), 
the number of degrees of freedom also vanishes. So, the holographic principle is not expected to hold in this case. 

\noindent However, if one takes Komar energy $(4 \pi/3) (3/\Lambda)^{3/2} \mid\rho_{\Lambda} + 3 p_{\Lambda}\mid$ (as in \cite{Padmanabhan2005}), one obtains $E = \cal{E}$ instead which, in this regard, is fully consistent with the holographic principle.
\\

\noindent From the energy, temperature and entropy of the horizon we obtain the Helmholtz free energy
\begin{equation}
F = {\cal{E}} - T\, S = \frac{1}{2} \sqrt{\frac{3}{\Lambda}},
  \label{F-energy1}  
\end{equation}
and, formally,  the partition function
\begin{equation}
F = - T \ln Z    \qquad \quad \Rightarrow  \; \; Z = \exp(-3 \pi/\Lambda). 
    \label{partition1}
\end{equation}
In doing so, we implicitly assume to be working with the canonical ensemble and that the gravitational field plays the role of the heat reservoir.
\\    \

\noindent It is immediately seen that $Z$ is related to the square of the scale factor, $ a^{2}(t) = g_{rr}$, and its first temporal derivative by
\begin{equation}
\ln Z = -  \frac{4 \pi}{\left[\frac{d}{dt} \left(\ln g_{rr}\right)\right]^{2}}. 
    \label{Z-grr}
\end{equation}
Thus, in this simple case, all thermodynamic information can be derived from the $g_{rr}$ term of the metric. 

\section{Power law universes}
\noindent Here we study whether the horizon of a flat Friedmann-Robertson-Walker (FRW) universe whose scale factor follows the power law $a(t) = t^{n}$  $(n>0$), fulfills the equipartition theorem. We shall consider first the possibility $n \leq 1$ (i.e., $\ddot{a} \leq 0$), and then the possibility $n >1$ (i.e., $\ddot{a} > 0$).
\\  

\noindent \underline{(i) Case $ 0 < n \leq 1$}

\noindent In this situation there is no event horizon, only Hubble horizon of radius equal to the inverse of the Hubble rate,  
$H = n/a^{1/n}$. The entropy of the horizon  is $S = \textstyle{1\over 4}{\cal{A}}  =  \frac{\pi}{n^{2}} a^{2/n}$. Taking into account that $\dot{H} = - n/a^{2/n}$ and that the temperature of dynamical horizons  \cite{Cai-Kim2005,Narayan2021}
\begin{equation}
T = \frac{H}{2 \pi} \left(1 + \frac{\dot{H}}{2 H^{2}}\right)
\label{dynamicaltemp}
\end{equation}
is in this case
\begin{equation}
T = \left(\frac{2n-1}{4 \pi} \right) a^{-1/n},
    \label{dynamicaltempn1}
\end{equation}
upon integrating Gibbs equation, we find the energy associated  to  the Hubble horizon
\begin{equation}
{\cal{E}} = \left(\frac{2n-1}{4 n^{2}} \right)\, a^{1/n}.
\label{upon-integration2}    
\end{equation}
Notice that this is positive for $n > 1/2$ only. Clearly, in the limiting case  $n = 1/2$ (corresponding to a radiation dominated universe) the unphysical result ${\cal{E}} = 0$ follows. However, this  is unrealistic because radiation always comes accompanied by matter whence, in reality, $n$ is somewhat greater than $1/2$. Likewise the case $0 < n <1/2$ is also unrealistic; in particular the sound speed in the fluid sourcing the gravitational field would exceed the speed of light.

\noindent Recalling that the number of degrees of freedom of the horizon is $4 \pi a^{2/n}/n^{2}$, 
the right hand side of equation (\ref{eq:eqp1}) coincides with the right hand side of (\ref{upon-integration2}) 
for $\zeta = 1/4$. It is intriguing that the equipartition theorem is satisfied in this case although the system 
is not at equilibrium, since $dS/da > 0$, and the horizon has a negative heat capacity.
\\  

\noindent Next, we study whether the holographic principle, in the sense that $E$ and ${\cal E}$ coincide with each other, is fulfilled.
\noindent Assuming the energy inside the volume enclosed by the horizon is given by $E = (4 \pi/3) \rho \, r_{H}^{3}$, with 
$\rho = 3H^{2}/(8 \pi)$, we get $E =\frac{a^{1/n}}{2n}$.  If we resort to $ E = (4 \pi/3) (\rho + p) r_{H}^{3} = - (\dot{H}/3) r_{H}^{3}$,
we find $E = a^{1/n}/(3n^{2})$. If Komar energy, $\mid (\rho+3p) (4 \pi/3) r_{H}^{3{}}\mid$, 
where $(\rho+3p) = -(3/4 \pi) (\ddot{a}/a)$, is used instead, it follows $E = [(n-1)/n^{2}] a^{1/n}$. Another way to determine $E$ is to integrate the expression for the flux of energy crossing inside the horizon, $ dE/dt =- ({\cal{A}}/4 \pi) \dot{H}$. It yields $E = a^{1/n}/n$.
\\ 

\noindent As is apparent, although all four expressions for $E$ are proportional to $a^{1/n}$, neither matches the right-hand side of (\ref{upon-integration2}). So, none of them fully complies with the holographic principle in the sense of above.

\noindent  Since the Helmholtz free energy, $F = {\cal{E}} - TS$, vanishes in this case, we have $Z = 1$.
\\

\noindent \underline{(ii) Case $n > 1$}

 \noindent In this case it exists an event horizon of radius $r_{H} = a^{1/n}/(n-1)$ and entropy \newline  $S = \pi a^{2/n}/(n-1)^{2}$. Realizing that the temperature of this dynamical horizon is \newline $T = (2n - 1) a^{-1/n}/(4 \pi)$, and resorting as before to Gibbs equation,  we obtain the energy associated to the horizon
\begin{equation}
{\cal{E}} = \frac{2n-1}{2n(n-1)^{2}}\, \int_{0}^{a}{x^{(1/n)-1} dx} = \frac{(2n-1)}{2(n-1)^{2}}\, a^{1/n}.
\label{horizonenergy(ii)}
\end{equation}

\noindent Clearly, the equipartition theorem, Eq. (\ref{eq:eqp1}), is fully satisfied for $\zeta  = 1/2$. 
\\

\noindent The energy inside the horizon, calculated via $(4 \pi/3) \, \rho \, r_{H}^{3}$ yields  \newline $n^{2} \, a^{1/n}/[2 (n-1)^{3}]$, and by means of Komar energy  $n \, a^{1/n}/(n-1)^{2}$. Although neither expression coincides with the right hand side of (\ref{horizonenergy(ii)}), it would not be fair to think that the holographic principle fails since in both instancess ($n \leq 1$ and $n > 1$), the energy associated to the horizon and the energy of the bulk are proportional to $a^{1/n}$. As it can be easily reasoned from Einstein equations, the fact that in all the cases considered in this section the energy of the bulk and the energy associated to the horizon are proportional to $a^{1/n}$ (i.e., to $t$) is simply because the time has dimensions of energy.
\\   

\noindent The partition function of the event horizon is readily  obtained as $Z = \exp \left[- \frac{\pi}{(n-1)^{2}} a^{2/n}\right]$.\\Recalling once more that $a^{2}$ is the $g_{rr}$ term of the FRW metric, it follows that (similarly to the De Sitter case) here, as
well as in case (i),  this term contains all thermodynamic information about the horizon.

\section{Pole-like universes}
\noindent In this case the scale factor reads  $a(t)= (t_{*} - t)^{-n}$,  where $n >1$ and $t_{*}$ is a constant satisfying 
$t_{*} \geq t$. Consequently,  this type of universe, proposed by Pollock and Singh \cite{Pollock-Singh1989}, expands faster than exponentially and leads to a ``big rip" scenario at late times \cite{Caldwell2003}. The radius of the event horizon $r_{H} = [n/(1+n)] H^{-1} = (t_{*} - t)/(1+n)$, decreases with expansion and so does its area,  $d{\cal A}/da < 0$, as well as its entropy.  A calculation parallel to those of the previous sections, via Gibbs equation, of the horizon  energy shows that it is negative,
\begin{equation}
{\cal{E}} = -\frac{1}{1+n} \int_{a_{i}}^{a} {dx/x} = - \frac{1}{1+n} \ln \left( \frac{a}{a_{i}}\right) \qquad        \qquad (t_{*}-t = a^{-1/n}), 
\label{pole-lieenergy1^3}
\end{equation}
\noindent and diverges as the scale factor approaches $a_{*}$. This result does not seem unreasonable in a ``big rip" scenario. Here $a_{i} = t_{*}^{-n}$ corresponds to the initial time, $t_{i} = 0$.
\\

\noindent The expression for $\cal{E}$ that corresponds to Eq. (\ref{eq:eqp1}) is $2 \zeta \, [(n+1) a^{1/n}]^{-1}$, whereby the theorem of equipartition of energy does not hold in this case. 
\\ 

\noindent This is consistent with the fact that the fields that drive super-acceleration, $\dot{H} > 0$ (phantom fields),  do not comply with the dominant energy condition and suffers from classical \cite{Dabrowski2015} and quantum \cite{Cline2004} instabilities. Moreover, since the area of the horizon decreases so does the volume enclosed by it, as well as the entropy of the horizon and that of matter and fields enclosed by the horizon. In view of this one might rush to  conclude that this kind of universes are not consistent with thermodynamics since they do not satisfy neither the equipartition theorem nor the holographic principle. However, based on quantum-mecachanical arguments, Gonz\'{a}lez-D\'{i}az et al. \cite{pedro2004} reasoned that phantom-dominated universes are also compatible with thermodynamics provided that negative temperatures are allowed.

\section{Comments and conclusions}
\noindent We studied  the theorem of equipartition of energy of statistical mechanics and the holographic principle in scenarios dominated by gravity, specifically in cosmological horizons. Likewise we found that, in various cases, from the metric of the system the partition function of the latter can be found. 

\noindent Our results can be summarized as follows: (i) The said theorem is fulfilled for the De Sitter horizon and the causal horizon of universes whose scale factor varies with time as $a = t^{n}$. (ii) In those cases, the energy conditions are satisfied and the parameter $\zeta$ takes either the $1/2$ value or $1/4$. The  possibility $\zeta = 1/2$ is intriguing because this value corresponds very generally to harmonic systems while cosmological horizons are not harmonic. (iii) Causal horizons of universes that fail to comply with the dominant energy condition do not fulfill the theorem because these systems violate the generalized second law of thermodynamics, as it is the case of pole-like universes studied in section fourth. (iv) The holographic principle, in the sense that the energy associated to the horizon exactly coincides with the Komar energy of the bulk, is complied by the De Sitter universe only. (v) However, in the case of power law universes the aforesaid principle approximately true because the energy of the bulk and that associated to the horizon have identical dependence on the scale factor, namely $a^{1/n}$. (vi) In some cases the term $g_{rr}$ of the metric of homogeneous and isotropic universes that comply with the second law is directly related to the partition function of the system, therefore it carries all thermodynamic information about the horizon. 
\\  \   

\noindent Before closing, we would like to emphasize the following. While the equipartition theorem is valid (as we have seen, in the examples of above) for gravitating systems that comply with the weak and dominant energy conditions, no proof exists thus far about its general validity. Note that the proof of that theorem in non-gravitational physics rests  on the assumption of thermal equilibrium between the system and some thermal bath; but in the cases here considered, the horizon is not in contact with any thermal bath. Nevertheless, one might naively think that the gravitational field may play the role of such bath. However, to ascribe a temperature to the gravitational field itself has no physical sense; somehow it is like ascribing a temperature to the electromagnetic field; formally it can be done but it lacks of any real meaning. So, a proof of the theorem for self-gravitating systems that satisfy the weak and dominant energy conditions would be most welcome.   In any case, the fact that causal horizons comply with the equipartition theorem of statistical mechanics (modulo the aforesaid  energy conditions are fulfilled) reinforces the connection between gravity and thermodynamics. A similar statement can be made about the fluctuations of the flux of energy that crosses the apparent horizon. These fluctuations behave as expected in non-gravitating thermodynamic systems \cite{mimoso2018}.  It remains to be seen if the equipartition theorem plays so an important role in gravitational physics and it does in non-gravitational statistical mechanics. Our guess is that it should play a similar role.
\\  

\noindent Needless to say that, in addition to that, it would be interesting to extend this study to the case of non-flat universes. We plan to undertake this in a 
future research.

\section*{Acknowledgments}
\noindent I am grateful to Narayan Banerjee for useful comments on a previous version of this manuscript and to the anonymous referee for advice.


\end{document}